\documentclass[10pt, a4paper, conference, compsocconf]{IEEEtran}
\usepackage{amsmath}
\usepackage{graphicx}
\usepackage{amsfonts}
\usepackage{amssymb}
\usepackage{alltt}
\usepackage{stmaryrd}
\usepackage{dsfont} 
\usepackage{color} 
\usepackage[standard]{ntheorem}
\usepackage{floatflt}
\usepackage{caption}
\usepackage[font=footnotesize]{subfig}
\usepackage{array}
\renewcommand{\theequation}{\arabic{equation}}
\usepackage[width=18.2cm,height=25.7cm]{geometry}

\ifCLASSINFOpdf
\else
\fi
\begin{document}

\title{A novel pseudo-random number generator based on discrete chaotic iterations}
\author{\IEEEauthorblockN{Qianxue Wang\IEEEauthorrefmark{1},
Christophe Guyeux\IEEEauthorrefmark{1} and
Jacques M. Bahi\IEEEauthorrefmark{1}}
\IEEEauthorblockA{\IEEEauthorrefmark{1}University of Franche-Comte\\
Computer Science Laboratory LIFC,
Belfort, France\\ Email:qianxue.wang@univ-fcomte.fr, christophe.guyeux@univ-fcomte.fr, jacques.bahi@univ-fcomte.fr}}

\maketitle

\begin{abstract}
Security of information transmitted through the Internet, against passive or active attacks is an international concern. The use of a chaos-based pseudo-random bit sequence to make it unrecognizable by an intruder, is a field of research in full expansion. This mask of useful information by modulation or encryption is a fundamental part of the TLS Internet exchange protocol. In this paper, a new method using discrete chaotic iterations to generate pseudo-random numbers is presented.
This pseudo-random number generator has successfully passed the NIST statistical test suite (NIST SP800-22). Security analysis shows its good characteristics. The application for secure image transmission through the Internet is proposed at the end of the paper.

\end{abstract}


\IEEEpeerreviewmaketitle

\section{Introduction}

Nowadays, the world is highly computerized and interconnected, this leads to a growing interest in the use of digital chaotic\footnote{In this document, chaos means Devaney's topological chaos~\cite{Dev89} which implies a deterministic but unpredictable system very sensitive to its initial conditions.} systems offering the possibility to reinforce the security of cryptographic algorithms, like those present in the Transport Layer Security protocol (TLS is an Internet exchange protocol).
The advantage of the use of chaotic dynamics for security problems lies in their unpredictability character and in the mathematical theory of chaos. This theory brings many qualitative and quantitative tools, namely  \linebreak ergodicity, entropy, expansivity and sensitive dependence to initial conditions, these tools allow the study of the randomness of the disorder generated by the considered system.

Most of these new applications use chaotic maps as pseudo-random number generators to obtain a binary stream, for example, for symmetric encryption.
Random number generators are essential in several fields like statistical studies, simulations (used for performance evaluations) or cryptography. They may be based on physical noise sources or on mathematical algorithms. However, in both cases, truly random numbers are not obtained because of data acquisition systems in the first case and machine precision in the second one. Instead, any real implementation actually produces a pseudo-random number generator (PRNG). Before using those generators in cryptographic applications, some strong requirements must be checked, for instance, they have to pass the up-to-date National Institute of Standards and Technology (NIST) statistical test suite \cite{ANDREW2008}, they should possess a long cycle length and a good entropy,~\emph{etc.} At the same time, the PRNG must also pass usual evaluations using traditional digital signal processing tools (autocorrelation function, cross-correlation function and fast Fourier transform).

The behaviors of chaotic dynamical systems are very similar to those of physical noise sources~\cite{Schuster1984}. Their sensitivity to initial conditions and their broadband spectrum make them good candidates to generate cryptographically secure PRNGs. Particularly, they have several basic properties that any good PRNG must possess: a long cycle length, strong randomness and entropy, speed, reproducibility,~\emph{etc.} However, chaotic dynamical systems are usually continuous and hence defined on the real numbers domain. The transformation from real numbers to integers may lead to the loss of the chaotic behavior. The conversion to integers needs a rigorous theoretical foundation.

\medskip

In this paper, a new chaotic pseudo-random bit generator is presented, which can also be used to obtain numbers uniformly distributed between 0 and 1. Indeed, these bits can be grouped $n$ by $n$, to obtain the floating part of $x \in [0,1]$ represented in binary numeral system. This generator is based on discrete chaotic iterations which satisfy Devaney's definition of chaos~\cite{guyeux09}. A rigorous  framework is introduced, where topological chaotic properties of the generator are shown. This generator successfully passes the whole NIST statistical tests. Moreover, because of its topological chaotic properties, this generator can be used for cryptographic applications.

\bigskip

The rest of this paper is organized in the following way. In Section \ref{Basic recalls}, some basic definitions concerning chaotic iterations and PRNGs are recalled. Section \ref{The novel generator based on discrete chaotic iterations} is devoted to the new generator which is based on discrete chaotic iterations,  all the design steps of this PRNG are described. In Section \ref{Statistical tests and Experiments} the results of  some experiments and statistical tests are given. In Section \ref{An application example of the proposed PRNG}, some application examples are proposed in the field of Internet secure exchanges. Some conclusions and future work end the paper.

\section{Basic recalls}
\label{Basic recalls}
This section is devoted to basic notations and terminologies in the fields of chaotic iterations, Devaney's chaos and pseudo-random number generators.

\subsection{Chaotic iterations}
\label{Chaotic iterations}
In the sequel $\llbracket 1;\mathsf{N} \rrbracket $ means $\{1,2,\hdots,N\}$, $s^{n}$  denotes the $n^{th}$ term of a sequence $s=(s^{1},s^{2},\hdots)$, $V_{i}$
denotes the $i^{th}$ component of a vector $V=(V_{1},V_{2},\hdots)$ and $f^{k}$
denotes the $k^{th}$ composition of a function $f$,
\begin{equation}
\begin{array}{r@{\;}l}
\ f^{k}=\underbrace{f\circ ...\circ f} \\
\ k\ \text{times}
\end{array}
\end{equation}

Let us consider a \emph{system} of a finite number $\mathsf{N}$ of \emph{%
cells}, so that each cell has a boolean \emph{state}. Then a sequence of
length $\mathsf{N}$ of boolean states of the cells corresponds to a
particular \emph{state of the system}. A sequence which elements belong in $%
\llbracket 1;\mathsf{N} \rrbracket $ is called a \emph{strategy}. The set of
all strategies is denoted by $\mathbb{S}.$

\begin{definition}
Let $S\in \mathbb{S}$. The \emph{shift} function is defined by $\sigma
:(S^{n})_{n\in \mathds{N}}\in \mathbb{S}\longrightarrow (S^{n+1})_{n\in %
\mathds{N}}\in \mathbb{S}$ and the \emph{initial function} $i$ is the map
which associates to a sequence, its first term: $i:(S^{n})_{n\in \mathds{N}%
}\in \mathbb{S}\longrightarrow S^{0}\in \llbracket1;\mathsf{N}\rrbracket$.
\end{definition}

\begin{definition}
The set $\mathds{B}$ denoting $\{0,1\}$, let $f:\mathds{B}^{\mathsf{N}%
}\longrightarrow \mathds{B}^{\mathsf{N}}$ be an iteration function and $S\in \mathbb{S}
$ be a chaotic strategy. Then, the so-called \emph{chaotic iterations} are defined by~\cite{Robert1986}

\begin{equation}
\begin{array}{l}
x^0\in \mathds{B}^{\mathsf{N}}, \\
\forall n\in \mathds{N}^{\ast },\forall i\in \llbracket1;\mathsf{N}\rrbracket%
,x_i^n=\left\{
\begin{array}{l}
x_i^{n-1} ~~~~~\text{if}~S^n\neq i \\
f(x^n)_{S^n} ~\text{if}~S^n=i.\end{array} \right. \end{array}\end{equation}
\end{definition}

In other words, at the $n^{th}$ iteration, only the $S^{n}-$th cell is
\textquotedblleft iterated\textquotedblright . Note that in a more general
formulation, $S^n$ can be a subset of components and $f(x^{n})_{S^{n}}$ can
be replaced by $f(x^{k})_{S^{n}}$, where $k\leqslant n$, describing for
example delays transmission (see \emph{e.g.}~\cite{Bahi2000}). For the
general definition of such chaotic iterations, see, e.g.~\cite{Robert1986}.

Chaotic iterations generate a set of vectors (boolean vector in this paper), they are defined by an initial state $x^{0}$, an iteration function $f$ and a chaotic strategy $S$.

\subsection{Devaney's chaotic dynamical systems}

Consider a metric space $(\mathcal{X},d)$ and a continuous function $f:%
\mathcal{X}\longrightarrow \mathcal{X}$. $f$ is said to be \emph{%
topologically transitive}, if for any pair of open sets $U,V\subset \mathcal{%
X}$, there exists $k>0$ such that $f^{k}(U)\cap V\neq \varnothing $. $(%
\mathcal{X},f)$ is said to be \emph{regular} if the set of periodic points
is dense in $\mathcal{X}$. $f$ has \emph{sensitive dependence on initial
conditions} if there exists $\delta >0$, such that, for any $x\in \mathcal{X}$
and any neighborhood $V$ of $x$, there exists $y\in V$ and $n\geqslant 0$
such that $|f^{n}(x)-f^{n}(y)|>\delta $. $\delta $ is called the \emph{%
constant of sensitivity} of $f$.

Quoting Devaney in \cite{Dev89}, a function $f:\mathcal{X}\longrightarrow
\mathcal{X}$ is said to be \emph{chaotic} on $\mathcal{X}$ if $(\mathcal{X}%
,f)$ is regular, topologically transitive and has sensitive dependence on
initial conditions.

When $f$ is chaotic, then the system $(\mathcal{X}, f)$ is chaotic and quoting Devaney it is unpredictable because of the sensitive dependence on initial conditions. It cannot be broken down or decomposed into two subsystems which do not interact because of topological transitivity. And in the midst of this random behavior, we nevertheless have an element of \linebreak regularity: fundamentally different behaviors are then possible and occurs with an unpredictably way.

The appendix gives the outline proof that chaotic iterations satisfy Devaney's topological chaos property. They can then be used to construct a new pseudo-random bit generator.

\subsection{Low-dimensional chaotic systems}

The dynamics of low dimension systems can be predicted using return map analysis or forecasting. Messages can thus be extracted from the chaos~\cite{Robilliard2006}. In addition, its randomness nature is deteriorated when a finite precision arithmetic is used. The chaotic properties are reduced: some \linebreak  severe problems such as short cycle length, non-ideal \linebreak distribution and high-correlation have been observed~\cite{HS2007}.

Therefore, it is required to merge two or more low-dimensional chaotic systems, to form a composite one~\cite{LiCQ2007}\cite{Zhang2005}\cite{LiSJ2001}\cite{Pareek2006}. With respect to this requirement, a new method based on discrete chaotic iterations is proposed in the next section.

\section{The novel generator based on discrete chaotic iterations}
\label{The novel generator based on discrete chaotic iterations}

The design of the new pseudo-random number generator based on discrete chaotic iterations, satisfying Devaney's chaos, is proposed and discussed. Detail operations of this approach are described in this section, while their performance will be presented in the next section.

\subsection{Chaotic iterations as pseudo-random generator}

The novel generator is designed by the following process.

Let $\mathsf{N} \in \mathds{N}^*, \mathsf{N} \geqslant 2$. Some chaotic iterations are done, which generate a sequence $\left(x^n\right)_{n\in\mathds{N}} \in \left(\mathds{B}^\mathsf{N}\right)^\mathds{N}$ of boolean vectors: the successive states of the iterated system. Some of those vectors are chaotically extracted and their components constitute our pseudo-random bit flow.

Chaotic iterations are realized as follows: initial state\linebreak $x^0 \in \mathds{B}^\mathsf{N}$ is a boolean vector taken as a seed, explained in Subsection \ref{algo seed} and chaotic strategy $\left(S^n\right)_{n\in\mathds{N}}\in \llbracket 1, \mathsf{N} \rrbracket^\mathds{N}$ is constructed from a logistic map $y$ (eq. \ref{suite logistique} in Subsection \ref{algo strategie}). Last, iterate function $f$  is the vectorial boolean negation
$$f_0:(x_1,...,x_\mathsf{N}) \in \mathds{B}^\mathsf{N} \longmapsto (\overline{x_1},...,\overline{x_\mathsf{N}}) \in \mathds{B}^\mathsf{N}.$$

To sum up, at each iteration, only $S^i$-th component of state $X^n$ is updated, as follows

\begin{equation}
x_i^n = \left\{\begin{array}{ll}x_i^{n-1} & \text{if } i \neq S^i, \\ \\ \overline{x_i^{n-1}} & \text{if } i = S^i. \\\end{array}\right.
\end{equation}

Finally, let $\mathcal{M}$ be a finite subset of $\mathds{N}^*$. Some $x^n$ are selected by a sequence $m^n$ as the pseudo-random bit sequence of our generator, where a sequence $(m^n)_{n \in \mathds{N}} \in \mathcal{M}^\mathds{N}$ is computed with $y$ (eq. \ref{mn} in Subsection \ref{algo m}). So, the generator returns the following values:
\begin{itemize}
\item the components of $x^{m^0}$,
\item following by the components of $x^{m^0+m^1}$,
\item following by the components of $x^{m^0+m^1+m^2}$,
\item \emph{etc.}
\end{itemize}

In other words, the generator returns the following bits:

\begin{small}
\noindent $x_1^{m_0}x_2^{m_0}x_3^{m_0}\hdots x_\mathsf{N}^{m_0}x_1^{m_0+m_1}x_2^{m_0+m_1}x_3^{m_0+m_1}\hdots x_\mathsf{N}^{m_0+m_1}$
$x_1^{m_0+m_1+m_2}x_2^{m_0+m_1+m_2}x_3^{m_0+m_1+m_2}...$
\end{small}

and its $k^{th}$ bit is $$\displaystyle{x_{k+1 \text{ (mod }\mathsf{N}\text{)}}^{\sum_{i=0}^{\lfloor k/\mathsf{N} \rfloor}m_i}}.$$

\begin{figure}[!t]
\centering
\includegraphics[width=2.5in]{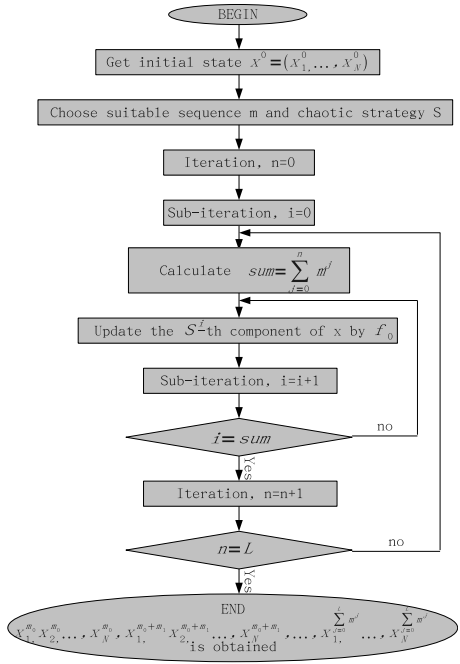}
\DeclareGraphicsExtensions.
\caption{Flow chart of chaotic strategy}
\label{Flow chart of chaotic strategy}
\end{figure}

The basic design steps of the novel generator are also presented in flow chart form in Figure ~\ref{Flow chart of chaotic strategy} ($N \cdot L$ is the length in bits of obtained sequence).

$\mathsf{N} = 5$ and $\mathcal{M} = \{4,5\}$ are adopted in the following subsections for easy understanding.

\subsection{Chaotic strategy}
\label{algo strategie}

Let $y^0 \in ]0;1[$ be a real number deduced  as a seed too (see Subsection \ref{algo seed}) and $y=(y^n)_{n \in \mathds{N}} \in [0,1]$ the logistic sequence defined as bellow

\begin{equation}
\label{suite logistique}
\forall n \in \mathds{N}, y^{n+1} = 4 y^n (1-y^n)
\end{equation}

Chaotic strategy is then the sequence $(S^n)_{n \in \mathds{N}} \in \llbracket 1 ; 5 \rrbracket^\mathds{N}$ equal to

\begin{equation}
\forall n \in \mathds{N}, S^{n} = \left( \lfloor 10^7 y^n \rfloor\right) \text{ mod } 5  + 1
\end{equation}

\subsection{Sequence $m$ of returned states}
\label{algo m}

Let us recall that $m^n$ is the number of iterations between the $n^{th}$ return of 5 pseudo-random bits and the following $n+1^{th}$ return. To define $(m^n)_{n\in \mathds{N}}$, the chaotic sequence of equation \ref{suite logistique} is used another time:

\begin{equation}
\label{mn}
\forall n \in \mathds{N}, m^{n} = \left\{\begin{array}{ll} 4 & \text{if } y^n < 0.5 \\ 5 & \text{if } y^n \geqslant 0.5 \\\end{array}\right.
\end{equation}

\subsection{Parameters of the generator}
\label{algo seed}

The initial state of the system $x^0$ and the first term $y^0$ of the logistic map are seeded by the current time in seconds since the Epoch, or a number that the user inputs.

Different ways are possible. For example, let us denote by $t$ the decimal part of the current time. So $x^0$ can be $t \text{ (mod 32)}$ written in binary digits ($2^5 = 32$ and the system is constituted by 5 bits) and $y^0 = t$.

\subsection{Illustration example}

In this example, the current time in seconds since the Epoch is 1237632934.484076. So, $t = 484076$, \linebreak $x^0 = t \text{ (mod 32)}$ in binary digits, \emph{i.e.} $x^0 = (1, 0, 1, 0, 0)$ and $y^0 = 0.484076$.

\medskip

Then
\begin{itemize}
\item y = 0.484076, 0.998985..., 0.004053..., 0.016146..., 0.063543.., 0.238022..., 0.725470..., 0.796651...
\item $m$ = 4, 5, 4, 4, 4, 4, 5, 5, 5, 5, 4, 5, 4,...
\item $S$ = 2, 4, 2, 2, 5, 1, 1, 5, 5, 3, 2, 3, 3,...
\end{itemize}

Chaotic iterations are made with initial state $x^0$, vectorial logical negation and strategy $S$, as shown in Table \ref{table application example}, and $m$ gives the states $x^n$ to return: $x^4, x^{4+5}, x^{4+5+4}, ...$.

\begin{tiny}
\begin{table*}[!t]
\renewcommand{\arraystretch}{1.3}
\caption{Application example}
\label{table application example}
\centering
  \begin{tabular}{|c|ccccc|cccccc|cccccc|}
    \hline
$m:$    &   &   & 4 &   &         &   &   & 5 &   &   &          &   &   & 4 &   &          & \\ \hline
$S$     & 2 & 4 & 2 & 2 &         & 5 & 1 & 1 & 5 & 5 &          & 3 & 2 & 3 & 3 &          & \\ \hline
$x^{0}$ &   &   &   &   & $x^{4}$ &   &   &   &   &   & $x^{9}$  &   &   &   &   & $x^{13}$ & \\
1       &   &   &   &   &
1       &   & $\xrightarrow{1} 0$ & $\xrightarrow{1} 1$ & & &
1      &   &   &   & &
1 & \\
0       & $\xrightarrow{2} 1$ & & $\xrightarrow{2} 0$ & $\xrightarrow{2} 1$ &
1       &   &   &   &   &  &
1       &   & $\xrightarrow{2} 0$ & & & 0 &\\
1       &   &   &   &   &
1       &   &   &   &   &  &
1       & $\xrightarrow{3} 0$ & & $\xrightarrow{3} 1$ & $\xrightarrow{3} 0$ &
0 &\\
0       &   & $\xrightarrow{4} 1$  &   & &
1       &   &   &   &   &  &
1       &   &   &   &   &
1 &\\
0       &   &   &   &   &
0       & $\xrightarrow{5} 1$ &   &  & $\xrightarrow{5} 0$ & $\xrightarrow{5} 1$ &
1      &   &   &   &   &
1 &\\
\hline
  \end{tabular}\\
\vspace{0.5cm}
Output: $x_1^{0}x_2^{0}x_3^{0}x_4^{0}x_5^{0}x_1^{4}x_2^{4}x_3^{4}x_4^{4}x_5^{4}x_1^{9}x_2^{9}x_3^{9}x_4^{9}$
$x_5^{9}x_1^{13}x_2^{13}x_3^{13}x_4^{13}x_5^{13}... = 10100111101111110011...$
\end{table*}
\end{tiny}

In this situation, the output of the generator is: 10100111101111110011...

\section{Statistical tests and Experiments}
\label{Statistical tests and Experiments}
The security of the new scheme is evaluated via both theoretical analysis and experiments.

\subsection{NIST statistical test suite}

\subsubsection{Presentation}

Among the numerous standard tests for pseudo-randomness, a convincing way to show the randomness of the produced sequences is to confront them to the NIST (National Institute of  Standards and Technology) Statistical Test: an up-to-date\footnote{A new version of the Statistical Test Suite (Version 2.0) has been released in August 25, 2008.} test suite by the Information Technology Laboratory (ITL).

The NIST test suite, SP 800-22, is a statistical package consisting of 15 tests. They were developed to test the randomness of (arbitrarily long) binary sequences produced by either hardware or software based cryptographic random or pseudo-random number generators. These tests focus on a variety of different types of non-randomness that could occur in a sequence.

\subsubsection{Interpretation of empirical results}

$\mathbb{P}$ is the tail probability that the chosen test statistic will assume values that are equal to or worse than the observed test statistic value when cosidering the null hypothesis.
For each statistical test, a set of $\mathbb{P}$s is produced from a set of sequences obtained by our generator (i.e., 100 sequences are generated and tested, hence 100 $\mathbb{P}$s are produced). The interpretation of empirical results can be conducted in any number of ways. In this paper, the examination of the distribution of $\mathbb{P}$s to check for uniformity ($\mathbb{P}_T$) is used.

The distribution of $\mathbb{P}$s is examined to ensure uniformity.

If $\mathbb{P}_T \geq 0.0001$, then the sequences can be considered to be uniformly distributed.

In our experiments, 100 sequences (s = 100), each with 1,000,000-bit long, are generated and tested. If the $\mathbb{P}_T$ of any test is smaller than 0.0001, the sequences are considered to be not good enough and the generating algorithm is not suitable for usage.

Table~\ref{The passing rate} shows $\mathbb{P}_T$ of the sequences based on discrete chaotic iterations using different schemes. If there are at least two statistical values in a test, the test is marked with an asterisk and the average value is computed to characterize the statistical values. Different schemes are using different lengths $\mathsf{N}$ of the iterated system and different sets $\mathcal{M}$ (range of $m^i$ which gives the states to return).

We can conclude from Table \ref{The passing rate} that the worst situation is Scheme 1: it just can be observed that 3 out of 15 of the tests are failed. However, if we find a right combination of $\mathsf{N}$ and $\mathcal{M}$ (Scheme 6) a noticeable improvement is observed, and all the tests are passed.

\begin{table*}[!t]
\renewcommand{\arraystretch}{1.3}
\caption{SP 800-22 test results ($\mathbb{P}_T$)}
\label{The passing rate}
\centering
  \begin{tabular}{|l||c|c|c|c|c|c|}
    \hline
Scheme & 1 & 2 & 3 & 4 & 5 & 6 \\ \hline
$\mathsf{N}$ (size of the system) & $8$ & $8$ & $8$ & $5$ & $5$ &$5$ \\ \hline
$\mathcal{M}$ & $\{1\}$ & $\{8\}$ & $\{1,..,8\}$ & $\{4,5\}$ & $\{9,10\}$ &$\{14,15\}$ \\ \hline\hline

Frequency (Monobit) Test &   0&  0.289667 &  0 &0.108791   &0.026948  &0.851383\\ \hline
Frequency Test within a Block (M=20000)  &  0 & 0  &0   &0.699313   &0.262249  &0.383827\\ \hline
Runs Test &  0 & 0.955835  &0.816537   &0.739918   &0.419021  &0.319084\\ \hline
Test for the Longest Run of Ones in a Block &  0 & 0  &0   &0.834308   & 0.616305 &0.137282 \\ \hline
Binary Matrix Rank Test &  0 &  0 &0.699313   &0.935716   &0.153763  &0.699313 \\ \hline
Discrete Fourier Transform (Spectral) Test &  0 & 0  &0   &0.162606   &0.798139  &0.129620 \\ \hline
Non-overlapping Template Matching Test* (m=9) & 0  & 0  & 0  &0.482340   &0.410039  &0.484733 \\ \hline
Overlapping Template Matching Test  (m=9) &  0 & 0  &0   &0.401199   &0.678686  &0.474986 \\ \hline
Maurer’s “Universal Statistical” Test  (L=7,Q=1280) &  0 & 0.075719  &0.080519   &0.102526   &0.455937  &0.096578 \\ \hline
 Linear Complexity Test (M=500) &  0.955835 &0.474986   & 0.051942  &0.023545   &0.637119  &0.419021 \\ \hline
Serial Test* (m=10) &  0 & 0  &0   &0.308152   &0.369959  &0.534272 \\ \hline
Approximate Entropy Test (m=10) & 0  & 0  &0   &0   &0  &0.991468 \\ \hline
Cumulative Sums (Cusum) Test* & 0  & 0.553415  &0   & 0.661814  &0.840655  & 0.755309 \\ \hline
Random Excursions Test* & 0.015102  & 0.45675  &0.194299   &0.293228   &0.335133  &0.654062 \\ \hline
Random Excursions Variant Test* &0.045440   &0.49615   &0.145418   &0.330716   &0.574089  &0.553885 \\ \hline
Success & 3/15 & 7/15 & 6/15 & 14/15 & 14/15&15/15\\ \hline
    \hline
  \end{tabular}
\end{table*}

\subsection{Experiment results}

The PRNG adopted in this section is Scheme 6 of Table~\ref{The passing rate}.

\begin{figure*}[!t]
\centering
\subfloat [The auto-correlation]{\includegraphics[width=2.5in]{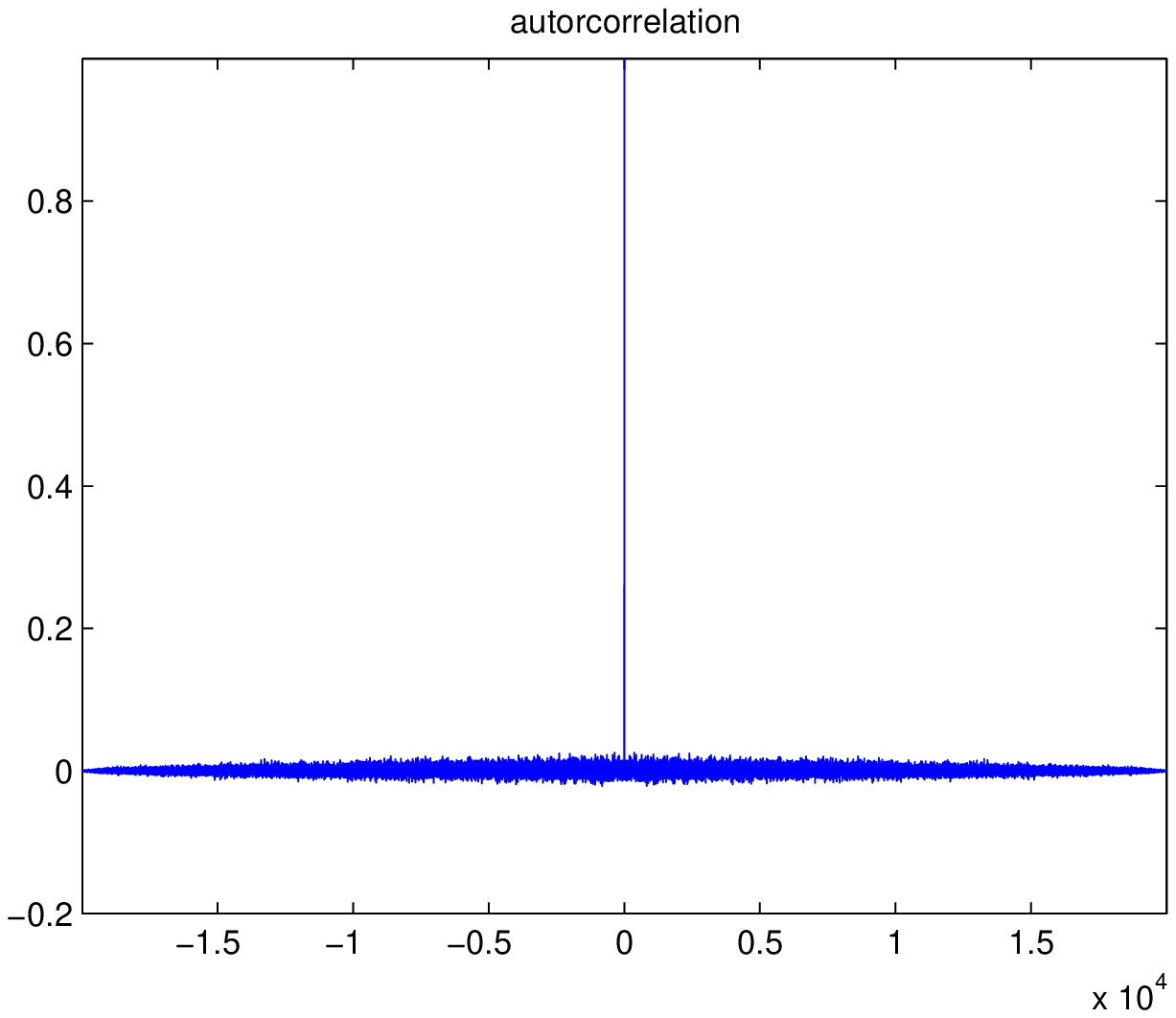}%
}
\hfil
\subfloat [The cross-correlation]{\includegraphics[width=2.5in]{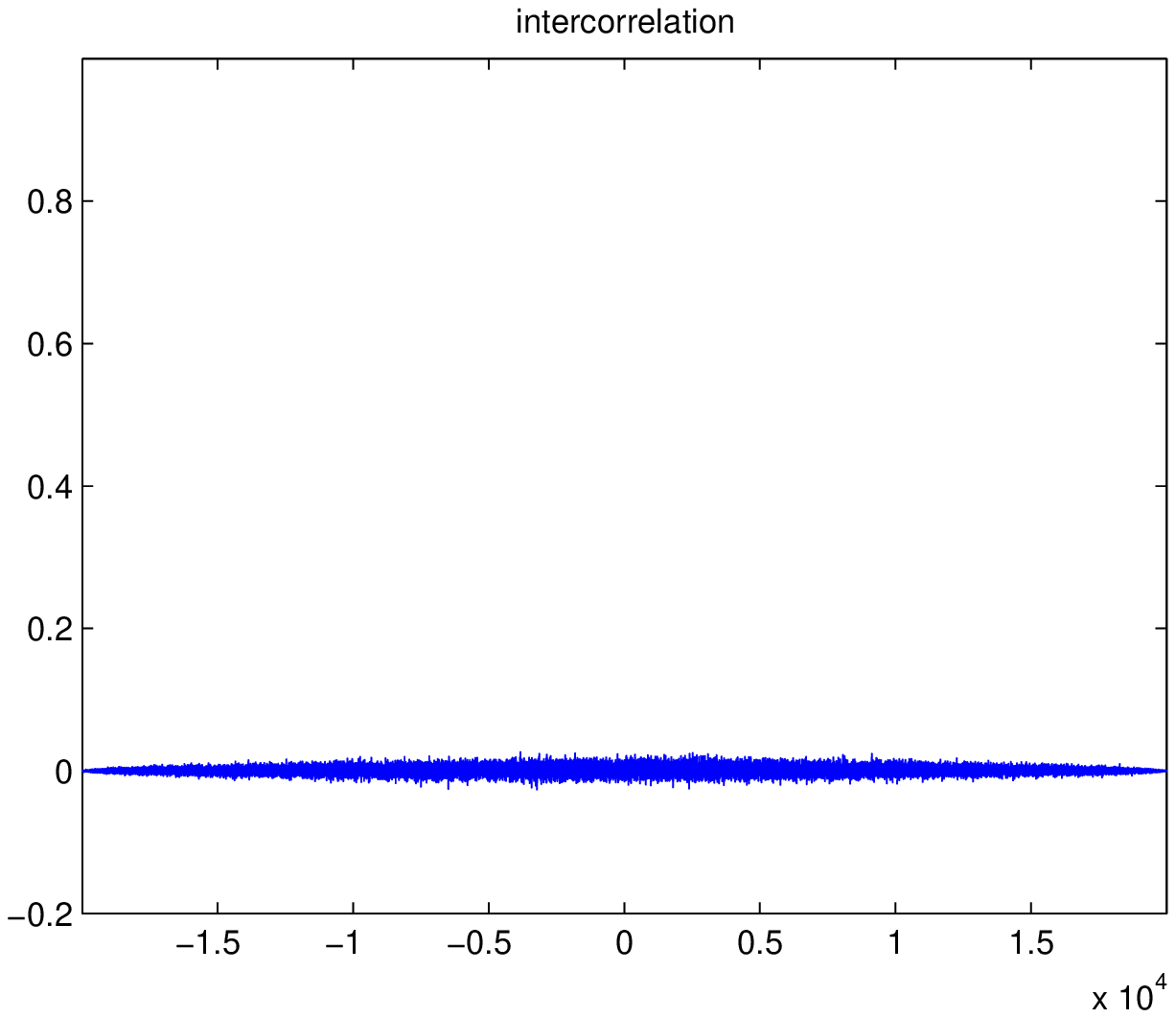}%
}
\caption{The auto-correlation and cross-correlation of the pseudo-random sequence}
\label{The auto-correlation &The cross-correlation of the pseudo-random sequence}
\end{figure*}

The auto-correlation and cross-correlation of the symbolic sequence are also given in Figure~\ref{The auto-correlation &The cross-correlation of the pseudo-random sequence}. It can be seen that this sequence has $\delta$-like auto-correlation which is required for a good PRNG. The sequences generated with different initial values will have zero cross-correlation due to the sensitive dependence on initial conditions.

\begin{figure}[!t]
\centering
\includegraphics[width=2.5in]{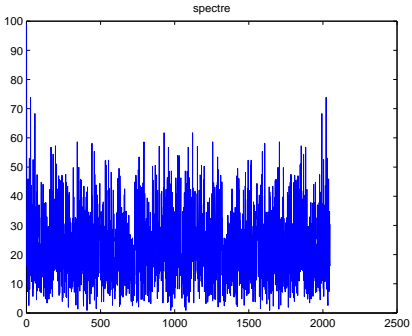}
\DeclareGraphicsExtensions.
\caption{The FFT of the pseudo-random sequence}
\label{The FFT of the pseudo-random sequence}
\end{figure}

The FFT of the sequence (Figure~\ref{The FFT of the pseudo-random sequence}) is performed and the corresponding power spectrum is computed. A complete flat power spectrum, with almost equal frequency contribution for all frequencies, is indicative of a total random serie.

\subsection{On the periodicity of chaotic orbit}
\label{Conclusions and Future Work}
\label{Experiments and statistical tests}

Suppose the system is realized in $k$-bit finite precision (under fixed-point arithmetic) and then digital chaotic iterations are constrained in a discrete space with $2^{k}$ elements, it is obvious that every chaotic orbit will eventually be periodic~\cite{Lishu2003}, i.e., finally go to a cycle with limited length not greater than $2^{k}$.

\begin{figure}[!t]
\centering
\includegraphics[width=2.5in]{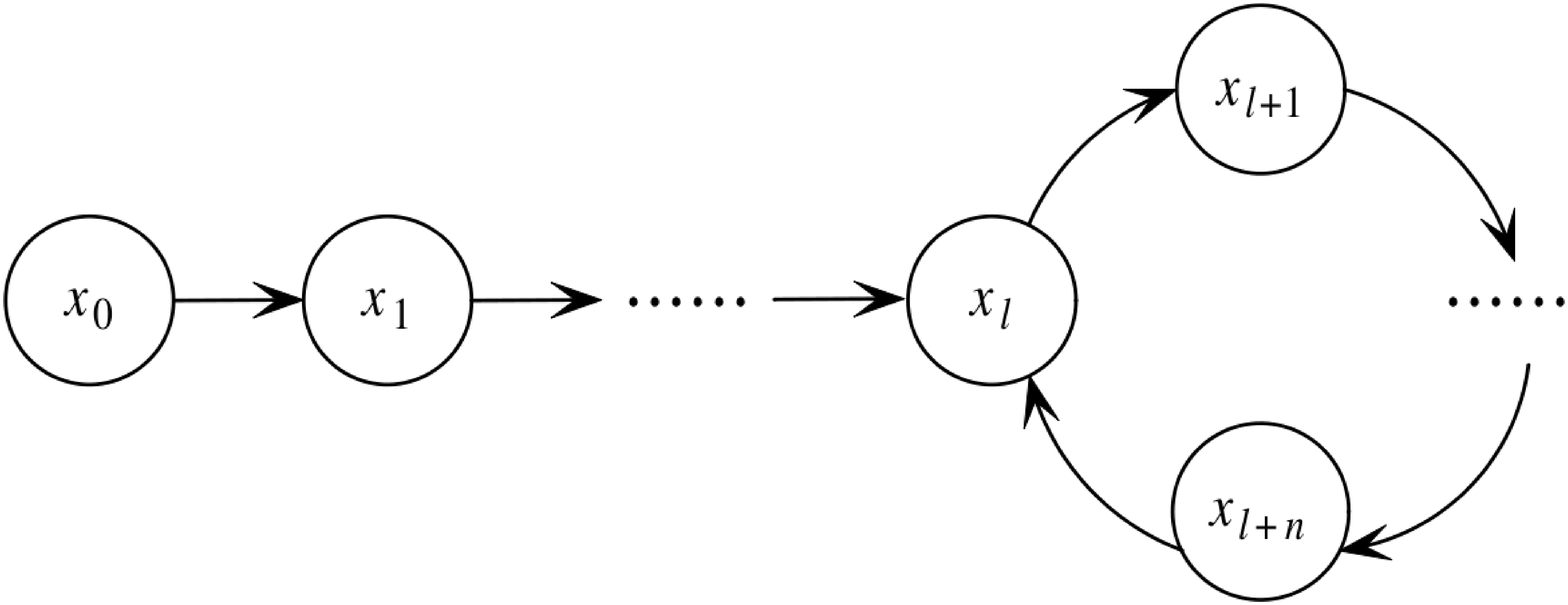}
\DeclareGraphicsExtensions.
\caption{A pseudo orbit of a digital chaotic system}
\label{A pseudo orbit of a digital chaotic system}
\end{figure}

The schematic view of a typical orbit of a digital chaotic system is shown in Figure~\ref{A pseudo orbit of a digital chaotic system}. Generally, each digital chaotic orbit includes two connected parts: $x^{0} , x^{1} , \dots, x^{l-1}$ and $ x^{l} , x^{l+1} , \dots , x^{l +n}$ , which are respectively called transient (branch) and cycle in this paper. Accordingly, $l$ and $n + 1$ are respectively called transient length and cycle period, and $l+n+1$ is called orbit length.

\begin{definition}
A sequence $X = (x^{ 1} , ..., x^{n} )$ is said cyclic if a subset of successive terms is repeated from a given rank, until the end of $X$.
\end{definition}

This novel generator based on discrete chaotic iterations generated by two pseudo-random sequences ($m$ and $S$) has a long cycle length. If the cycle period of $m$ and $S$ is  $n_{m}$ and $n_{S}$, in an ideal situation, the cycle period of the new sequence is $n_{m}\cdot n_{S}\cdot2$ (because $\bar{\bar{x}}=x$).

\begin{example}

m ($n_{m}=2$): \textbf{12}121212121212121212121212...

$S$ ($n_{S}=4$): \textbf{1 23 4} 12 3 41 2 34 1 23 4 12 3 41 2 34 1 23 4...

X($n_{X}=2 \cdot 4 \cdot 2 = 16$): \textbf{0000(0) 1000(8) 1110(14) 1111(15) 0011(3) 0001(1) 1000(8) 1100(12) 1111(15) 0111(7) 0001(1) 0000(0) 1100(12) 1110(14) 0111(7) 0011(3)} 0000(0) 1000(8) 1110(14) 1111(15) 0011(3) 0001(1) 1000(8) 1100(12) 1111(15) 0111(7) 0001(1) 0000(0) 1100(12) 1110(14) 0111(7) 0011(3)...
\end{example}

\section{An application example of the proposed PRNG}
\label{An application example of the proposed PRNG}
Cryptographically secure PRNGs are fundamental tools to communicate securely through the Internet.

\begin{figure}[!t]
\centering
\subfloat [The original image]{\includegraphics[scale=0.4]{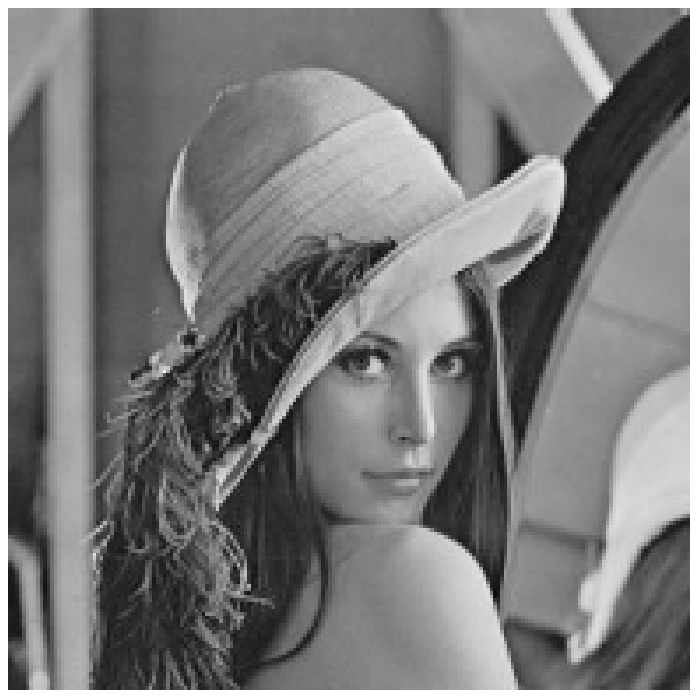}%
}
\hfil
\subfloat [The histogram of original image]{\includegraphics[scale=0.39]{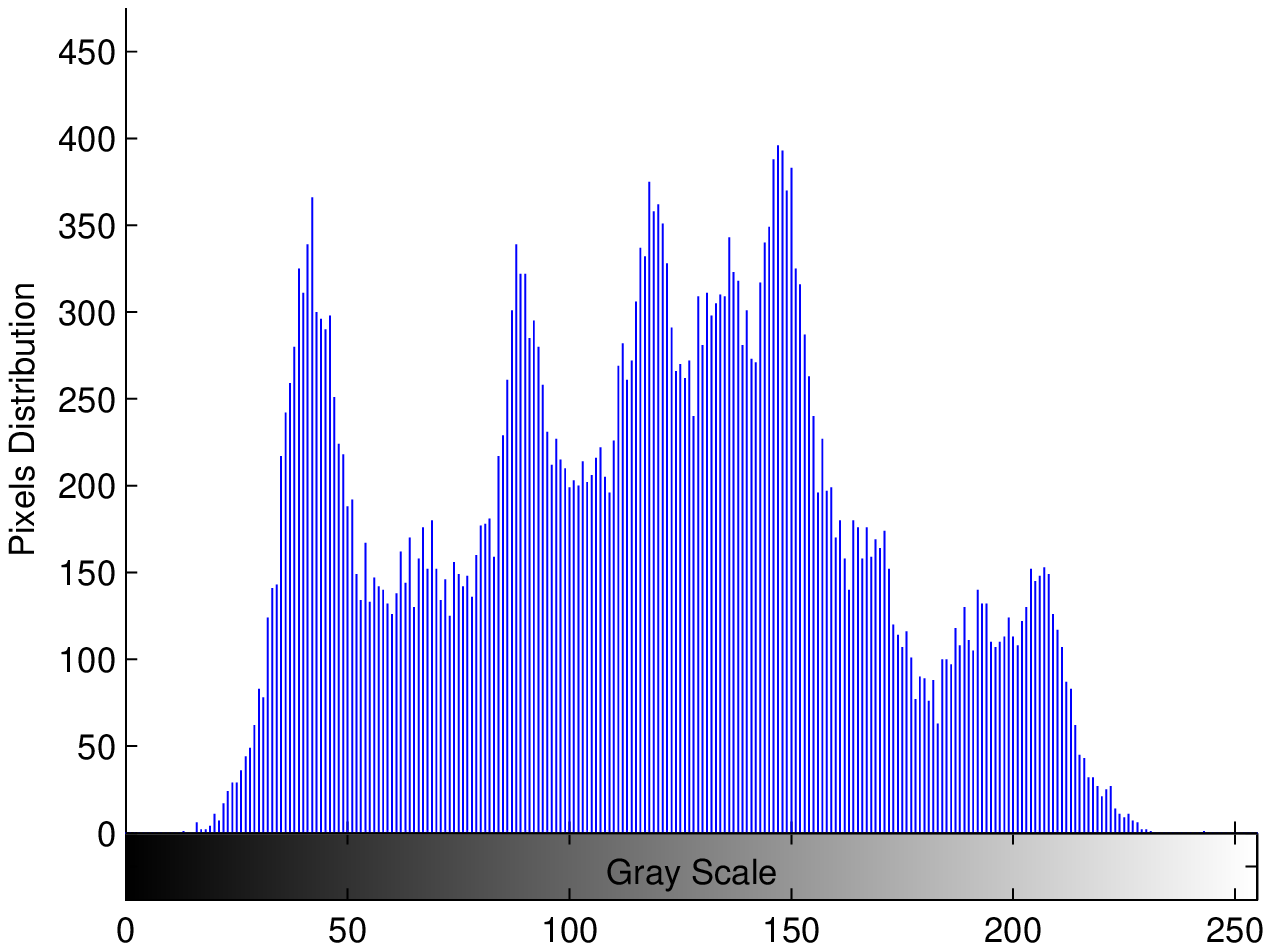}%
}
\caption{Distribution of original image}
\label{Distribution of original image}
\end{figure}

\begin{figure}[!t]
\centering
\subfloat [The encrypted image]{\includegraphics[scale=0.4]{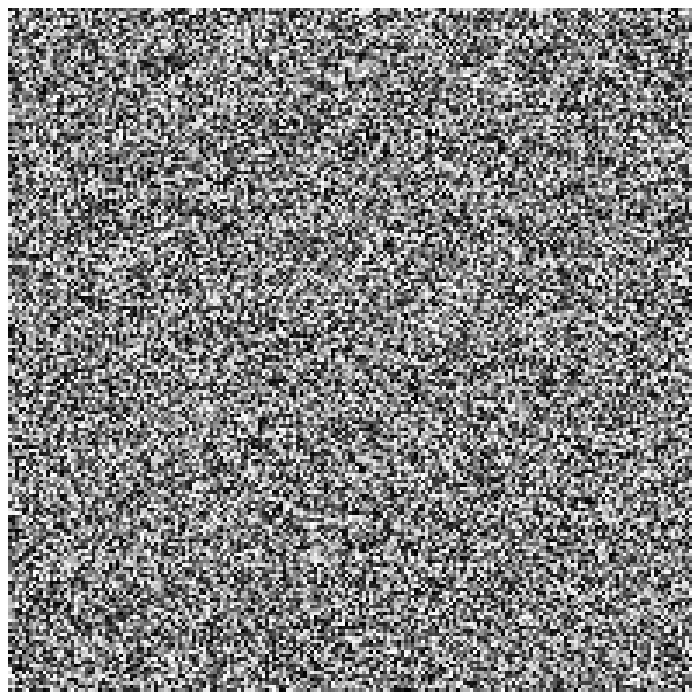}%
}
\hfil
\subfloat [The histogram of encrypted image]{\includegraphics[scale=0.39]{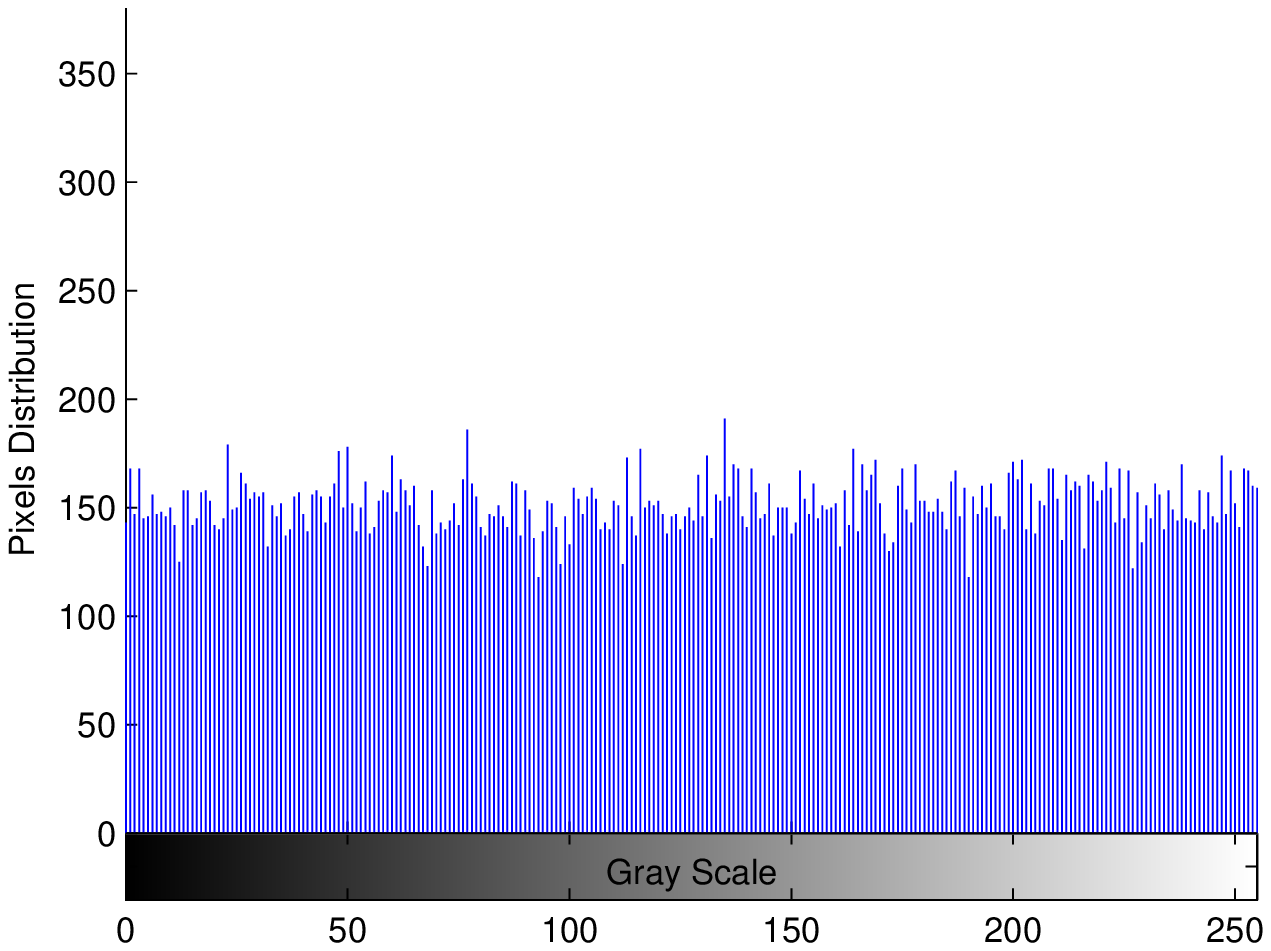}%
}
\caption{Distribution of encrypted image}
\label{Distribution of encrypted image}
\end{figure}

For example, in order to guarantee security of image transmission, the previous pseudo-random sequence can be used to encrypt the digital image (one-time pad encryption). The original image and the encrypted image are shown in Figures~\ref{Distribution of original image}(a) and~\ref{Distribution of encrypted image}(a). Figures~\ref{Distribution of original image}(b) and~\ref{Distribution of encrypted image}(b) depict the histograms. It can be seen that the distribution of the encrypted image is very close to the uniform distribution, which can well protect the information of the image to withstand the statistical attack.

TLS protocol is another example in which cryptographically secure PRNGs are needed, during the generation of private key for symmetric cypher. This generation requires a high quality of the randomness for the PRNG.

\bigskip

The generator presented in this paper has passed the whole NIST800-22 statistical test suite, so it can reasonably be considered as a possibly usable PRNG. We believe that this generator can also be used for cryptographic applications, because of its topological chaos quality. Indeed, it is proved in ~\cite{guyeux09} that Devaney's chaos property is satisfied by the discrete chaotic iterations: they are regular, transitive and sensitive to initial conditions.

Because of transitivity, the discrete dynamical system cannot be decomposed: the behavior of the system cannot be reduced to the study of one of its parts. As a consequence, the knowledge of a part of the private key (or the encrypted image) cannot help an hypothetical attacker to guess the whole key (image). Moreover, the sensitiveness conducts to the fact that, even if the attacker tries anyway to decrypt the cypher message by attempting to complete the part in his possession, he cannot succeed.

Last the regularity participates to an increase of the randomness of our generator and conducts to the impossibility of the prediction of its future evolution. Two very similar sequences can have completely different behaviors after some iterations, the first can quickly enter into a cycle whereas the second can follow a more divergent trajectory. Thus, two different seeds generate completely different keys.

\section{Conclusions and future work}
\label{Conclusions and Future Work}

In this paper, a novel pseudo-random generator based on discrete chaotic iterations is proposed. Different schemes are used to generate this chaotic sequence. A particular scheme (Scheme 6)  offers a sufficiently secure randomness for cryptographic applications. The proposed PRNG is based on a rigorous framework. In addition, a detailed statistical analysis concerning the numbers produced by this method is given. These experimental results lead us to conclude that our generator is a very good and reliable PRNG and that chaotic iterations can be used in computer science security field\cite{guyeux09}.

In future work, different random sequences will be used in place of logistic map, the influence of $\mathsf{N}$ and the range $\mathcal{M}$ of $m^{j}$ for the output sequence will be explored and other iteration functions will be studied. New applications in computer science field will be proposed, specially in the security and cryptography domains.

\bibliographystyle{plain}
\bibliography{A_novel_generator}

\section*{APPENDIX}

In this appendix we give outline proofs of the properties on which our pseudo-random number generator is based.

Denote by $\delta $ the \emph{discrete boolean metric},\linebreak $\delta
(x,y)=0\Leftrightarrow x=y.$ Given a function $f$, define the function $%
F_{f}:$ $\llbracket1;\mathsf{N}\rrbracket\times \mathds{B}^{\mathsf{N}%
}\longrightarrow \mathds{B}^{\mathsf{N}}$ such that $$F_{f}(k,E)=\left(
E_{j}.\delta (k,j)+f(E)_{k}.\overline{\delta (k,j)}\right) _{j\in \llbracket%
1;\mathsf{N}\rrbracket},$$ where + and . are the boolean addition and product operations.

Consider the phase space: $\mathcal{X}=\llbracket1;\mathsf{N}\rrbracket^{%
\mathds{N}}\times \mathds{B}^{\mathsf{N}}$ and the map $$G_{f}\left( S,E\right) =\left( \sigma
(S),F_{f}(i(S),E)\right) ,$$ then the chaotic iterations defined in (\ref%
{Chaotic iterations}) can be described by the following iterations
\[
\left\{
\begin{array}{l}
X^{0}\in \mathcal{X} \\
X^{k+1}=G_{f}(X^{k}).%
\end{array}%
\right.
\]

Let us define a new distance between two points $(S,E),(\check{S},\check{E})\in
\mathcal{X}$ by $$d((S,E);(\check{S},\check{E}))=d_{e}(E,\check{E})+d_{s}(S,%
\check{S}),$$ where
\begin{itemize}
\item $\displaystyle{d_{e}(E,\check{E})}=\displaystyle{%
\sum_{k=1}^{\mathsf{N}}\delta (E_{k},\check{E}_{k})} \in \llbracket 0 ; \mathsf{N} \rrbracket$ \\
\item $\displaystyle{%
d_{s}(S,\check{S})}=\displaystyle{\dfrac{9}{\mathsf{N}}\sum_{k=1}^{\infty }%
\dfrac{|S^{k}-\check{S}^{k}|}{10^{k}}} \in [0 ; 1].$
\end{itemize}

\medskip

It is then proved in \cite{guyeux09} by using the sequential continuity that

\begin{proposition}
\label{continuite} $G_f$ is a continuous function on $(\mathcal{X},d)$.
\end{proposition}

Then, the vectorial negation $f_{0}(x_{1},%
\hdots,x_{\mathsf{N}})=(\overline{x_{1}},\hdots,\overline{x_{\mathsf{N}}})$ satisfies  the three conditions for Devaney's chaos, namely, regularity and transitivity and sensitivity  in the metric space $(\mathcal{X},d)$. This leads to the following result.

\begin{proposition}
$G_{f_0}$ is a chaotic map on $(\mathcal{X},d)$ in the sense of Devaney.
\end{proposition}

\end{document}